\documentclass[conference]{IEEEtran}
\pdfoutput=1
\usepackage{comment}

\usepackage{cite}
\usepackage{amsmath,amssymb,amsfonts}
\usepackage{graphicx}
\usepackage{textcomp}

\usepackage{cite}
\usepackage{amsmath,amssymb,amsfonts}
\usepackage{algorithmic}
\usepackage{graphicx}
\usepackage{epstopdf}
\usepackage{subfigure}
\usepackage{stfloats}
 \graphicspath{{\figures}}
\usepackage{textcomp}
\usepackage{xcolor}
\usepackage{verbatim}
\usepackage{makecell}
\usepackage{booktabs}
\usepackage{CJK}
\usepackage{multirow}

\usepackage{algorithm}
\usepackage{algorithmic}

\makeatletter
\newcommand{\removelatexerror}{\let\@latex@error\@gobble}
\makeatother

\IEEEoverridecommandlockouts   
\begin{document}
\begin{CJK}{UTF8}{gbsn}

\title{\LARGE  Sum-Rate Maximization for Active RIS-Aided Downlink RSMA System}    

\author{Xinhao Li\IEEEauthorrefmark{1}, Tao Wang\IEEEauthorrefmark{1}, Haonan Tong\IEEEauthorrefmark{1}, Zhaohui Yang\IEEEauthorrefmark{2}, Yijie Mao\IEEEauthorrefmark{3}, 
Changchuan Yin\IEEEauthorrefmark{1}\\
 
 \small \IEEEauthorrefmark{1}  
 
Beijing Key Laboratory of Network System Architecture and Convergence,\\
Beijing University of Posts and Telecommunications, Beijing, China\\
\IEEEauthorrefmark{2}
College of Information Science and Electronic Engineering, Zhejiang University, Hangzhou, China\\

\IEEEauthorrefmark{3}
School of Information Science and Technology, ShanghaiTech University, Shanghai,  China\\
Emails: \{lxnh, taowang, hntong, ccyin\}@bupt.edu.cn, yang\_zhaohui@zju.edu.cn,
maoyj@shanghaitech.edu.cn\\
}

\maketitle

\begin{abstract}
In this paper, the problem of sum-rate
maximization for an active reconfigurable intelligent surface
(RIS) assisted downlink rate-splitting multiple access (RSMA) transmission system is studied.
 In the considered model, the active RIS is deployed
to overcome severe power attenuation, which is caused by the cumulative product of RIS incidence path loss and the reflection path loss. Since the active RIS can adjust both the phase and 
the amplitude of the incident signal simultaneously, the RIS control scheme requires delicate design to improve  RSMA communication performance. 
To address this issue, a sum-rate maximization problem is formulated to jointly optimize the beamforming vectors,  rate allocation vector, and RIS precoding matrix. To solve this non-convex sum-rate maximization problem, an iterative algorithm based on fractional programming (FP) and quadratic constraint quadratic programming (QCQP) is proposed. In particular, the proposed algorithm firstly decomposes the original problem into two subproblems, namely, 1) beamforming and rate allocation optimization and 2) active RIS precoding optimization. The corresponding variables of the two subproblems are  optimized through sequential convex approximation (SCA) and  block coordinate descent (BCD), respectively.   Numerical results show that the proposed active RIS-aided RSMA system could increase the sum-rate by up to 45\%
 over the conventional passive RIS-aided RSMA system with the same energy consumption.
\end{abstract}
~\\

\section{Introduction}


The future  networks will be confronted with complex wireless environments due to increasing wireless devices and 
emerging applications, which makes adjusting the wireless channel conditions vital to improve the communication performance\cite{ris6g}. To enable the wireless channel condition modification, reconfigurable intelligent surface (RIS), a planar surface comprised of a large number of tuneable elements, is proposed,  which can adjust the phase of the reflecting electromagnetic signals\cite{RISEE}. 
 Benefiting from the high array gain and low power consumption, RIS plays an important role on overcoming blockages and enhancing the spectrum efficiency of wireless networks  \cite{riseeeee}. However, most of existing works have ignored an essential characteristic of the passive RIS: The ``multiplicative fading" of the cascade transmitter-RIS-user link, which seriously restraints the performance of passive RIS\cite{activeris}.

   To overcome the high path-loss imposed
by the ``multiplicative fading" effect,  
the authors in \cite{activeris} designed a new type of RIS named by active
RIS. Active RIS can adjust the phase shift and amplify
the amplitude of the incident signals with low-cost negative resistance components that includes tunnel diode and negative impedance converter, etc.
    Inspired by \cite{activeris}, the prior works \cite{activeEEE,secure,zhuqi}  focused on exploiting the
advantages of the active RIS. In particular, it has been proved in \cite{activeEEE}  that the active RIS is more energy-efficient than the passive RIS as well as the amplify-and-forward  relay schemes. The authors in \cite{secure} investigated the secrecy performance of an active RIS-aided wireless communication system. Moreover, the authors in \cite{zhuqi} proposed a sub-connected active RIS architecture to reduce the power consumption of active RIS. However, all the aforementioned works \cite{activeris,activeEEE,secure,zhuqi} optimized the RIS system based on the orthogonal multiple access (OMA) or simply treated multi-user interference as noise, which  did
not fully utilize the multiplexing gain introduced by multiple antennas and therefore limits the channel capacity.
    
Non-orthogonal multiple access (NOMA) serves multiple users in the same time-frequency block by separating users in the power or code domain, which can achieve higher spectral efficiency than OMA\cite{activenoma}. 
However, using NOMA scheme, one user has to fully decode the interference from other users, leading to inefficient use of multiple antennas\cite{Ref3}. To address this issue, rate-spitting multiple access (RSMA) is introduced  \cite{Ref2},  where the data messages are splitted and encoded into one common and several private streams at the transmitter and are further decoded by successive interference cancellation (SIC) at the receivers.
The authors in \cite{Ref5,EERSMA}  also
proved the spectral and energy efficiency enhancement of RIS-aided
RSMA compared to RIS-aided NOMA scheme.
The main advantage of RSMA is that it can
flexibly manage the interference by allowing the interference
to be partially decoded and partially treated as noise. The
work in \cite{Ref7}  showed that RSMA is more spectrally efficient
and robust to channel state information (CSI)  impairments  than NOMA in a wide range of user deployments scenario. However, all the aforementioned work do not consider the use of active RIS to address the ``multiplicative fading" effect introduced by the passive RIS.

    The main contribution of this paper is a novel system that integrates  the active RIS into RSMA system. To our best knowledge, this is the first work that 
analyzes and optimizes the sum-rate of an active RIS-aided RSMA system. Our key contributions include:


    \begin{itemize}
    \item An active RIS-aided RSMA downlink transmission system model is developed, where the RIS adjusts both the
phase and the amplitude of the signal. A sum-rate maximization problem is formulated where the
transmit beamforming vectors, the rate allocation vector, 
and the active RIS precoding matrix are jointly optimized. 
    \item  To solve the non-convexity of the formulated problem,
the original problem is decomposed into two subproblems and the original problem is solved through iteratively optimizing these two subproblems.  In particular, the  beamforming and rate allocation subproblem is solved through  sequential convex approximation (SCA). And the optimal  RIS precoding matrix is obtained by fractional
progragmming (FP) and quadratic constraint quadratic programming (QCQP).

        \item Numerical analysis is conducted and simulation results show that the proposed algorithm  could converge efficiently. The active RIS-aided RSMA system  could achieve higher sum-rate than  the passive RIS-aided RSMA system,
with the same energy consumption.
    \end{itemize}
 
 The rest of the paper is organized as follows. Section \uppercase\expandafter{\romannumeral2} presents the  system model and problem formulation. Section \uppercase\expandafter{\romannumeral3} the algorithm design, while Section \uppercase\expandafter{\romannumeral4} presents the numerical results to evaluate the performance of active RIS-aided RSMA system. Finally, we conclude the paper in Section \uppercase\expandafter{\romannumeral5}.
    
\vspace{1em}
\section{System Model and Problem Formulation}

\subsection{System Model}
  We consider an active RIS-aided downlink multiuser multiple-input single-output (MU-MISO) RSMA system as shown in Fig. 1, where an $N_{\rm T}$-antenna base station (BS) serves $K$ single-antenna users simultaneously with the aid of an $L$-element active RIS.  Let $\mathcal{K}=\{1, \cdots, K\}$ denotes the set of users. The precoding matrix $\boldsymbol{\Psi}\in\mathbb{C}^{L\times L}$ of the active RIS can be written as $\mathbf{\Psi}=\mathbf{P} \mathbf{\Theta}$, where $\mathbf{\Theta} := \operatorname{diag}\left(e^{j \theta_{1}}, \cdots, e^{j \theta_{L}}\right)\in \mathbb{C}^{L\times L}$ is the phase-shift diagonal matrix, and $\mathbf{P}:=\operatorname{diag}\left(p_1, \cdots, p_L\right) \in \mathbb{R}_{+}^{L \times L}$ is the amplification factors matrix of the active RIS.

According to RSMA scheme, the transmit signal from the BS, $\mathbf{x}\in\mathbb{C}^{N_{\rm T}\times1}$, contains a common signal and $K$ private signals for the $K$ users, which is given by
 \vspace{0.5em}
        \begin{equation}
        \mathbf{x}= \underbrace{\mathbf{w}_0x_0}_\text{Common signal}+\underbrace{\sum_{k=1}^K{\mathbf{w}_kx_k}}_\text{Private signals},
     \end{equation}
where $x_0$, $x_k\in \mathbb{C}$ are normalized transmitted symbols and  $\mathbf{w}_0$, $\mathbf{w}_k\in\mathbb{C}^{N_{\rm T}\times1}$ are the corresponding common and the private beamforming vectors of user $k$, respectively.
   
    Let $\mathbf{g}_{k}\in   \mathbb{C}^{N_T\times1}$, $\mathbf{G}\in   \mathbb{C}^{L\times N_T}$, and $\mathbf{f}_{k}\in \mathbb{C}^{L\times1}$ denote the channel  between BS and user $k$, BS and the RIS, RIS and  user $k$, respectively.  
    	\begin{figure}[t]
	\includegraphics[scale = 0.75]{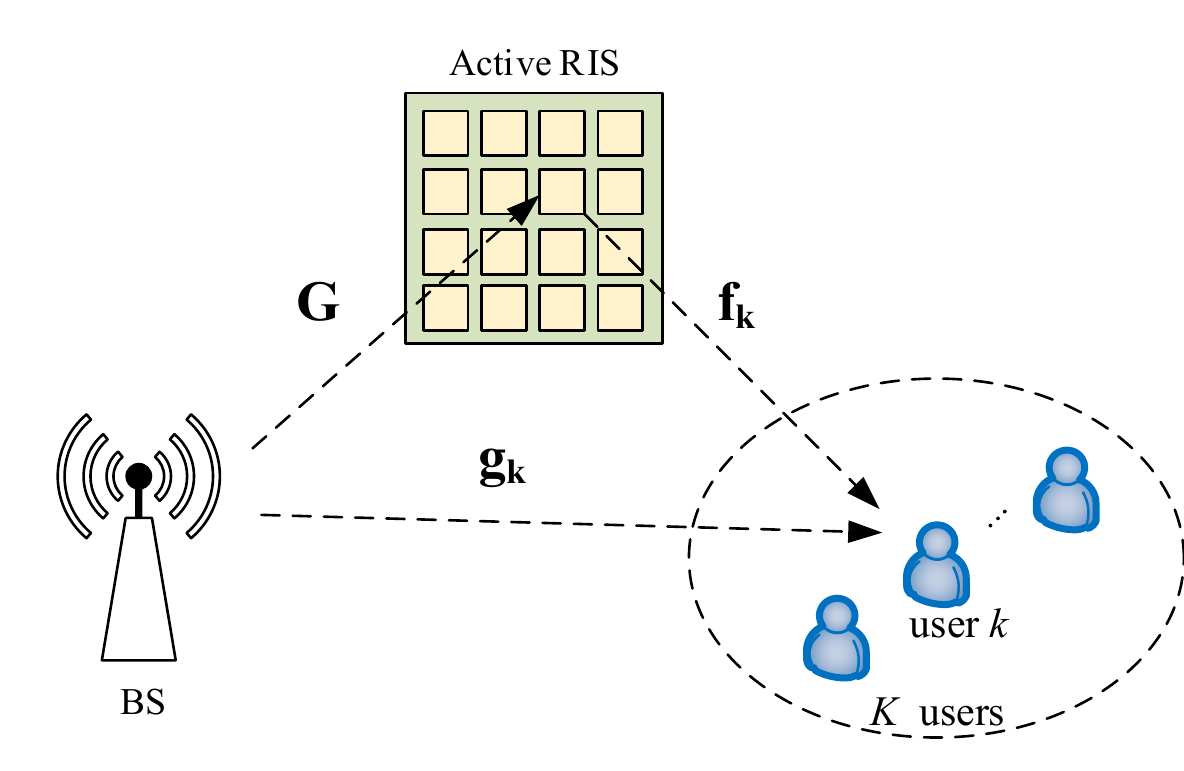}
	\caption{An active RIS-aided downlink MU-MISO RSMA system.}
 \label{fig:system}
 \vspace{-1em}
	\end{figure}
    Besides, we assume that perfect CSI at the transmitter is available. The equivalent channel from the BS to user $k$ can be denoted as
     \vspace{0.5em}
    \begin{equation}
        \mathbf{h}_{k}^{H}=\mathbf{g}_{ k}^{H}+\mathbf{f}_{ k}^{H} \boldsymbol{\Psi} \mathbf{G}, \quad\forall k \in \mathcal{K}.
    \end{equation}
The received signal $y_k$ of user $k$ is given as
\begin{equation}
    y_{k}=\mathbf{h}_{k}^{H}\mathbf{x}+\mathbf{f}_k^H \boldsymbol{\Psi} \boldsymbol{z}+n_{k},\quad\forall k \in \mathcal{K},
\end{equation}
 \vspace{0.5em}where $\boldsymbol{z} \sim \mathcal{C} \mathcal{N}\left(\boldsymbol{0}_L, \sigma_z^2 \boldsymbol{I}_L\right)$ and $n_k \sim \mathcal{C} \mathcal{N}\left(0, \sigma_k^2\right)$ denote the dynamic noise of the active RIS with power $\sigma_z^2$ and the additive white Gaussian noise (AWGN) with power $\sigma_k^2$, respectively.

Hence, the signal-to-interference-plus-noise ratio (SINR) of the common signal and the private signals at the user $k$ can be represented as
\begin{equation}
    \gamma_{c, k}=\frac{\left|\mathbf{h}_{k}^{H} \mathbf{w}_{\mathrm{0}}\right|^{2}}{\sum_{i=1}^{K}\left|\mathbf{h}_{k}^{H} \mathbf{w}_{i}\right|^{2}+\left\|\mathbf{f}_{k}^H \boldsymbol{\Psi}\right\|^2 \sigma_z^2+\sigma_{k}^{2}},\forall k\in \mathcal{K},
\end{equation}
\vspace{-0.5em}
\begin{equation}
      \gamma_{p, k}=\frac{\left|\mathbf{h}_{k}^{H} \mathbf{w}_{{k}}\right|^{2}}{\sum_{ i=1, i\neq k}^{K}\left|\mathbf{h}_{k}^{H} \mathbf{w}_{i}\right|^{2}+\left\|\mathbf{f}_{k}^H \boldsymbol{\Psi}\right\|^2 \sigma_z^2+\sigma_{k}^{2}},\forall k \in \mathcal{K},
\end{equation}
respectively.The achievable rate in bps/Hz of user $k$ to decode the common signal and its desired private signal are respectively expressed as
\begin{equation}
    R_{c, k}=\log _{2}\left(1+\gamma_{c, k}\right) \text { and } R_{p,k}=\log _{2}\left(1+\gamma_{p,k}\right), \forall k \in \mathcal{K},
\end{equation}
respectively. To ensure that the common signal can be successfully decoded by all  users, the achievable rate of the common signal should not exceed the channel capacity, which is given by: $\sum_{i=1}^K{r_{c,i}} \leq R_{c,k}$, $\forall k\in \mathcal{K}$, where  $\mathbf{r}_{c}=\left(r_{c,1}, \cdots, r_{c,k}\right)$ is the common rate allocation vector.

\subsection{Problem Formulation}
   Our aim 
is to maximize the user sum-rate, which is the sum of the common signal rate $\mathbf{r}_ c$ and achievable private signal rate. Particularly, we jointly design the transmit beamforming vectors $\mathbf{w}=\left[\mathbf{w}_{0}, \mathbf{w}_{1}, \cdots,  \mathbf{w}_{ K} \right]$ at the BS, rate allocation vector $\mathbf{r}_c$ ,and active RIS precoding matrix $\boldsymbol{ \Psi}$. Mathematically, the sum-rate maximization problem is formulated as 
\vspace{-0.5em}
\begin{align}
   \mathcal{P}_0: \max_{\mathbf{r}_{\mathrm{c}}, \mathbf{w},\boldsymbol{ \Psi}}& \quad\sum_{k=1}^{K}\left(r_{c,k}+R_{p,k}\right)\tag{7a}\\
    \text { s.t. }  &\sum_{i=1}^{K}r_{c,i} \leq R_{c, k}, \quad\forall k \in \mathcal{K},\tag{7b}\\  
& r_{c,k}+R_{p,k}\geq R_{k}^{\min},\quad\forall k \in \mathcal{K},\tag{7c}\\
&\sum_{k=0}^{K}\left\|\mathbf{w}_{k}\right\|^{2}\leq P_{\rm BS }^{\rm max},\tag{7d}\\
&\sum_{k=0}^{K}\left\|\mathbf{\Psi}\mathbf{\mathbf{G}w}_{k}\right\|^{2}+\|\mathbf{\Psi}\|^{2} \sigma_{z}^{2} \leq P_{\mathrm{A}}^{\max },\tag{7e}\\
     &r_{c,k}\geq 0,\tag{7f}
\end{align}
where constraint (7b) is imposed to ensure all users can successfully decode the common signal. Constraint (7c) is imposed to guarantee the required minimum rate by all users. (7d) and (7e) are BS power and active RIS power constraints, respectively.  Note that, problem $\mathcal{P}_0$ is non-convex, which is hard to directly solve. To tackle the non-convexity, we propose an algorithm that adopts the block coordinate descent (BCD) method to decompose the original problem into two subproblems in the next section. 
\vspace{-0.5em}
\section{Algorithm Design}
In this section, we propose an iterative algorithm to solve the problem $\mathcal{P}_0$.  We decompose $\mathcal{P}_0$ into two subproblems via BCD method, one subproblem for joint optimization of the transmit beamforming  $\mathbf{w}$ and the common rate  $\mathbf{r}_{c}$   with a given active RIS precoding matrix $\mathbf{\Psi}$, and the other subproblem is the optimization of the active RIS precoding matrix $\mathbf{\Psi}$  with given $\mathbf{r}_{c}$ and $\mathbf{w}$. 
\vspace{-0.5em}
\subsection{Joint Beamforming and Rate Allocation Optimization}
With the given active RIS precoding matrix $\mathbf{\Psi}$,  we attempt to deal with the non-convex
constraints in (7) via SCA method in this subsection. We introduce slack auxiliary variables $\boldsymbol{\xi}:=\left[\xi_{1}, \cdots, \xi_{K}\right]^{T} \in \mathbb{R}_{+}^{K}$,  $ \boldsymbol{\kappa}:=\left[\kappa_{1}, \cdots, \kappa_{K}\right]^{T} \in \mathbb{R}_{+}^{K}$ and rewrite the  original problem $\mathcal{P}_0$ as follows
\begin{align}
   \mathcal{P}_1: \max_{\mathbf{r}_{\mathrm{c}}, \mathbf{w},\mathbf{\xi},\mathbf{\kappa}}& \quad\sum_{k=1}^{K}\left(r_{c,k}+\log_{2}\left(1+\xi_k\right)\right)\tag{8a}\\
    \text { s.t. } &\sum_{i=1}^{K} r_{c, i} \leq \log_{2}\left(1+\kappa_k\right), \quad\forall k \in \mathcal{K},\tag{8b}\\  
     & \kappa_k\leq \frac{\left|{\mathbf{h}}_{k}^{H} \boldsymbol{w}_{0}\right|^{2}}{\sum_{i=1}^{K}\left|{\mathbf{h}}_{k}^{H} \boldsymbol{w}_{i}\right|^{2}+\sigma^{2}},\quad\forall k \in \mathcal{K},\tag{8c}\\
&\sum_{k=0}^{K}\left\|\mathbf{w}_{k}\right\|^{2}\leq P_{\rm BS }^{\rm max},\tag{8d}\\
   & \left\|\mathbf{\Psi}\mathbf{\mathbf{G}}\right\|^{2} \mathbf{w}^{H} \mathbf{w}\leq P_{\mathrm{A}}^{\max}-\|\mathbf{\Psi}\|^{2} \sigma_{z}^{2},\tag{8e}\\ 
    &\xi_k\leq \frac{\left|{\mathbf{h}}_{k}^{H} \mathbf{w}_{k}\right|^{2}}{\sum_{i=1, i \neq k}^{K}\left|{\mathbf{h}}_{k}^{H} \mathbf{w}_{i}\right|^{2}+\sigma^{2}},\quad\forall k \in \mathcal{K},\tag{8f}\\
    &\xi_k\geq 2^{R_{k}^{\rm min}-r_{c,k}}-1 ,\quad\forall k \in \mathcal{K},\tag{8g} \\
     &r_{c,k}\geq 0\tag{8h},
\end{align}
where $\sigma^2=\left\|\mathbf{f}_{k}^H \boldsymbol{\Psi}\right\|^2 \sigma_z^2+\sigma_{k}^{2}$. 

To deal with the non-convexity of constraint (8f), we next introduce an auxiliary variable $\beta_k\in \mathbb{R}_{+}$ and rewrite constraint (8f) as follows 
\begin{equation}
    \left|{\mathbf{h}}_{k}^{H} \mathbf{w}_{k}\right|^{2} \geq \beta_{k} \xi_{k},\quad\forall k \in \mathcal{K}\tag{9},
\end{equation}
and 
\begin{equation}
   \sum_{i=1, i \neq k}^{K}\left|{\mathbf{h}}_{k}^{H} \mathbf{w}_{i}\right|^{2}+\sigma^{2} \leq \beta_{k} ,\quad\forall k \in \mathcal{K}\tag{10}.
\end{equation}

Note that the term  $\mathbf{h}_{k}^{H} \mathbf{w}_{k}$ in constraint (9) can be expressed as a real number through an arbitrary rotation to beamforming vectors $\mathbf{w}_{k}$. Thus, constraint (9) is equivalently transformed to $\mathcal{R}\left({\mathbf{h}}_{k}^{H} \mathbf{w}_{k}\right) \geq \sqrt{\beta_{k} \xi_{k}}$. By adopting the first-order Taylor approximations, constraint (9) is equivalently transformed to
\begin{equation}\tag{11}
\begin{aligned}
\mathcal{R}\left({\mathbf{h}}_{k}^{H} \mathbf{w}_{k}\right) \geq & \sqrt{\beta_{k}^{(t-1)} \xi_{k}^{(t-1)}}+\frac{1}{2} \sqrt{\frac{\beta_{k}^{(t-1)}}{\xi_{k}^{(t-1)}}}\left(\xi_{k}-\xi_{k}^{(t-1)}\right) \\
&+\frac{1}{2} \sqrt{\frac{\xi_{k}^{(t-1)}}{\beta_{k}^{(t-1)}}\left(\beta_{k}-\beta_{k}^{(t-1)}\right)},\quad\forall k \in \mathcal{K},
\end{aligned}
\end{equation}
where the  superscript ($t-1$) represents that the corresponding variables take values at ($t-1$)-th iteration. Then we introduce an auxiliary variable $\lambda_k \in \mathbb{R}_{+}$, constraints (8c) is equivalently transformed to
\begin{equation}
    \left|\mathbf{h}_{k}^{H} \mathbf{w}_{0}\right|^{2} \geq \lambda_{k} \kappa_{k}=\frac{1}{4}\left(\left(\lambda_{k}+\kappa_{k}\right)^{2}-\left(\lambda_{k}-\kappa_{k}\right)^{2}\right)\tag{12},
\end{equation}
\begin{equation}
    \sum_{i=1}^{K}\left|\mathbf{h}_{k}^{H} \mathbf{w}_{i}\right|^{2}+\sigma^{2} \leq \lambda_{k},\quad\forall k \in \mathcal{K}\tag{13}.
\end{equation}

Note that (12) is non-convex, therefore, we adopt the difference-of-convex (DC) approximation and (12) is equivalently transformed to
\begin{equation}\tag{14}
\begin{aligned}
& 2 \mathcal{R}\left(\mathbf{h}_{k}^{H} \mathbf{w}_{0}^{(t-1)} {\mathbf{h}}_{k}^{H} \mathbf{w}_{0}\right)-\left|{\mathbf{h}}_{k}^{H} \mathbf{w}_{0}^{(t-1)}\right|^{2} \\
\geq & \frac{1}{4}\left(\left(\lambda_{k}+\kappa_{k}\right)^{2}-\left(\lambda_{k}^{(t-1)}-\kappa_{k}^{(t-1)}\right)\left(\lambda_{k}-\kappa_{k}\right)\right.\\
&\left.+\left(\lambda_{k}^{(t-1)}-\kappa_{k}^{(t-1)}\right)^{2}\right),\quad\forall k \in \mathcal{K}.
\end{aligned}
\end{equation}
With the above approximations, problem $\mathcal{P}_1$ can be transformed by  SCA method  as follows
\vspace{-0.5em}
\begin{align}
   \mathcal{P}_2:&\min_{\mathbf{w}, \mathbf{r_c}, \boldsymbol{\xi}, \boldsymbol{\beta}, \boldsymbol{\kappa},\boldsymbol{\lambda}} \quad-{\sum_{k=1}^{K}\left(r_{c,k}+ \log _{2}\left(1+\xi_{k}\right)\right)}\tag{15}\\
  \text { s.t. } &\quad(8\mathrm{b}), (8\mathrm{ d}), (8\mathrm{e}), (8\mathrm{g}), (8\mathrm{h}), (10), (11), (13), (14). \notag
\end{align}
Problem $\mathcal{P}_2$ is  convex, thus, the optimal  beamforming vectors ${\mathbf{w}^{*}_{m}}$ and the optimal rate allocation vector  ${\mathbf{r}^{*}_{c}}$ can be obtained by  CVX toolbox \cite{cvx}.

\subsection{Active RIS Precoding Optimization}
 Given the transmit beamforming 
 vectors $\mathbf{w}$ and the rate allocation $\mathbf{r}_{c}$, we attempt to adopt FP transform to deal with the non-convex object function in (7), and transform the problem $\mathcal{P}_0$ to a standard QCQP problem. Consider the characteristic of active RIS precoding
matrix, we denote $\psi_n=p_ne^{j \theta_{n}}$, $\forall n\in\mathcal{L}$, and $\boldsymbol{\psi}=\left[\psi_{1}, \cdots, \psi_{L}\right]^{\mathrm{T}}$ as the vectorized  $\boldsymbol{\Psi}$.  Specifically, through introducing auxiliary variable $\boldsymbol{\tau}: \triangleq\left[\tau_{1}, \tau_{2}, \cdots, \tau_{K}\right]^{T} \in \mathbb{R}_{+}^{K} $, we adopt the \textit{Lagrangian Dual Transform}\cite{fp}, the objective function in problem $\mathcal{P}_0$ can be equivalently reformulated as 
\begin{multline}\tag{16}
f_{1}\left( \boldsymbol{\psi}, \boldsymbol{\tau  }\right) \triangleq \sum_{k=1}^{K}\left [ \log _{2}\left(1+\tau _{k}\right)-\tau _{k}\right] \\
+ \sum_{k=1}^{K} \frac{\left(1+\tau_{k}\right)\left|\mathbf{h}_{k}^{H} \mathbf{w}_{k}\right|^{2}}{\sum_{i=1}^{K}\left|\mathbf{h}_{k}^{H} \mathbf{w}_{i}\right|^{2}+\left\|\mathbf{f}_ k^{H} \mathbf{\Psi}\right\|^{2} \sigma_{z}^{2}+\sigma_{k}^{2}},
\end{multline}
where $\boldsymbol{\tau} $ has the optimal solution as 
\begin{equation}\tag{17}
    \tau_{k}^{\star}=\frac{\left|\mathbf{h}_{k}^{H} \mathbf{w}_{k}\right|^{2}}{\sum_{i=1, i \neq k}^{K}\left|\mathbf{h}_{k}^{H} \mathbf{w}_{i}\right|^{2}+\left\|\mathbf{f}_{k}^{H} \mathbf{\Psi}\right\|^{2} \sigma_{z}^{2}+\sigma_{k}^{2}}, \forall k.
\end{equation}
Note that the last term in (16) is a sum of multiple fractions, we adopt \textit{Quadratic Transform} \cite{fp} on the fractional term and introduce the auxiliary variable $\boldsymbol{\upsilon}: \triangleq\left[\upsilon_{1}, \upsilon_{2}, \cdots, \upsilon_{K}\right]^{T} \in \mathbb{R}_{+}^{K} $, and (10) is further equivalently transformed to 
\begin{multline}\tag{18}
f_{2}\left(\boldsymbol{\psi}, \boldsymbol{\tau}, \boldsymbol{\upsilon }\right) \\
\triangleq \sum_{k=1}^{K}\left[\log _{2}\left(1+\tau_{k}\right)-\tau_{k}+2 \sqrt{1+\tau_{k}} \Re\left\{\upsilon _{k}^{*} \mathbf{h}_{k}^{H} \mathbf{w}_{k}\right\}\right. \\
\left.-\left|\upsilon _{k}\right|^{2}\left(\sum_{i=1}^{K}\left|\mathbf{h}_{k}^{H} \mathbf{w}_{i}\right|^{2}+\left\|\mathbf{f}_{ k}^{H} \boldsymbol{\Psi}\right\|^{2} \sigma_{z}^{2}+\sigma_{k}^{2}\right)\right],
\end{multline}
where the $\boldsymbol{\upsilon} $ has the optimal solution as 
\begin{equation}\tag{19}
    \upsilon _{k}^{\star}=\frac{\sqrt{1+\tau  _{k}} \mathbf{h}_{k}^{H} \mathbf{w}_{k}}{\sum_{i=1}^{K}\left|\mathbf{h}_{k}^{H} \mathbf{w}_{i}\right|^{2}+\left\|\mathbf{f}_{ k}^{H} \boldsymbol\Psi\right\|^{2} \sigma_{z}^{2}+\sigma_{k}^{2}}, \forall k .
\end{equation}
After the above transformation, problem $\mathcal{P}_0$ can be reformulated as
\begin{align}\max_{\boldsymbol{\psi},\boldsymbol{\tau},\boldsymbol{\upsilon}}&\quad f_{2}\left(\boldsymbol{\psi},  \boldsymbol{\tau}, \boldsymbol{\upsilon }\right)\tag{20a}\\
     \text{s.t.} \quad
     & \sum_{i=1}^{K}r_{c,i} \leq R_{c, k}, \quad\forall k \in \mathcal{K},\tag{20b}\\ 
     & r_{c,k}+R_{p,k}\geq R_{k}^{\min},\quad\forall k \in \mathcal{K},\tag{20c} \\
     & \sum_{k=0}^{K}\left\|\mathbf{\Psi}\mathbf{\mathbf{G}w}_{k}\right\|^{2}+\|\mathbf{\Psi}\|^{2} \sigma_{z}^{2} \leq P_{\mathrm{A}}^{\max }.\tag{20d}
 \end{align}
To deal with the non-convexity of constraints in problem (20),  we 
introduce the following notations:
\begin{equation}\tag{21}
      \mathbf{A}_{1p}=\mathbf{g}_{k}\mathbf{ w}_{k} \mathbf{w}_{k}^{{H}} \mathbf{g}_{k}^{H}, \quad \mathbf{A}_{1c}=\mathbf{g}_{k}\mathbf{ w}_{0} \mathbf{w}_{0}^{{H}} \mathbf{g}_{k}^{H}, 
\end{equation}
\begin{equation}\tag{22}
\mathbf{B}_{1p}=\operatorname{diag}\left(\mathbf{f}_{k}^{ {H}}\right) \mathbf{G w}_{k} \mathbf{w}_{k}^{ {H}} \mathbf{G}^{ {H}} \operatorname{diag}\left(\mathbf{f}_{k}\right), 
\end{equation}
\begin{equation}\tag{23}
\mathbf{B}_{1c}=\operatorname{diag}\left(\mathbf{f}_{k}^{ {H}}\right) \mathbf{G w}_{0} \mathbf{w}_{0}^{ {H}} \mathbf{G}^{ {H}} \operatorname{diag}\left(\mathbf{f}_{k}\right), 
\end{equation}
\begin{equation}\tag{24}
\mathbf{C}_{1p}=\operatorname{diag}\left(\mathbf{f}_{k}^{ {H}}\right) \mathbf{G w}_{k} \mathbf{w}_{k}^{ {H}} \mathbf{g}_{k}, \quad \mathbf{C}_{1c}=\operatorname{diag}\left(\mathbf{f}_{k}^{ {H}}\right) \mathbf{G w}_{0} \mathbf{w}_{0}^{ {H}} \mathbf{g}_{k}.
\end{equation}
Thus, the numerator of SINR  $\left|\mathbf{h}_{k}^{H} \mathbf{w}_{0}\right|^{2} $ in constraint (20b)  and $\left|\mathbf{h}_{k}^{H} \mathbf{w}_{k}\right|^{2} $  in constraint (20c) can be respectively rewritten as 
\begin{equation}\tag{25}
    \left|\mathbf{h}_{k}^{H} \mathbf{w}_{0}\right|^{2}= \mathbf{A}_{1c}+\boldsymbol{\psi}^{  {H}} \mathbf{B}_{1c}\boldsymbol{\psi}+\Re\left\{{\boldsymbol{\psi}^{  {H}}\mathbf{C}_{1c}}\right\},
    \end{equation}
and
\begin{equation}\tag{26}
    \left|\mathbf{h}_{k}^{H} \mathbf{w}_{k}\right|^{2}= \mathbf{A}_{1p}+\boldsymbol{\psi}^{  {H}} \mathbf{B}_{1p}\boldsymbol{\psi}+\Re\left\{{\boldsymbol{\psi}^{  {H}}\mathbf{C}_{1p}}\right\}.
\end{equation}
Similarly, to handle the non-convexity of the denominators of constraints (20b) and (20c), we introduce the following definitions 

\begin{equation}\tag{27}
\mathbf{D}=\sum_{k=1}^{K} \operatorname{diag}\left(\mathbf{f}_{k}^{ {H}}\right) \operatorname{diag}\left(\mathbf{f}_{k}\right) \sigma_{v}^{2},
\end{equation}
\begin{equation}\tag{28}
      \mathbf{E}=\mathbf{g}_{k}\sum_{j=1, j\neq k}^{K}\mathbf{w}_{j} \mathbf{w}_{j}^{ {H}}\mathbf{g}_{k}^{H}+\sigma^{2}_k, \quad
      \mathbf{A}_{2p}=\mathbf{A}_{1p}+\mathbf{E}, 
\end{equation}
\begin{equation}\tag{29}
\mathbf{F}=\operatorname{diag}\left(\mathbf{f}_{k}^{ {H}}\right)  \mathbf{G }\sum_{j\neq k}^{K}\mathbf{w}_{j} \mathbf{w}_{j}^{ {H}}\mathbf{G}^{ {H}} \operatorname{diag}\left(\mathbf{f}_{k}\right),
\end{equation}
\begin{equation}\tag{30}
\mathbf{B}_{2p}=\mathbf{B}_{1p}+\mathbf{F}, 
\end{equation}
\begin{equation}\tag{31}
\mathbf{J}= \operatorname{diag}\left(\mathbf{f}_{k}^{ {H}}\right) \mathbf{G }\sum_{j=1,j\neq k}^{K}\mathbf{w}_{j} \mathbf{w}_{j}^{ {H}} \mathbf{g}_{k},\quad
\mathbf{C}_{2p}= \mathbf{C}_{1p}+\mathbf{J}.
\end{equation}
Then the denominators of SINR  constraints (20b) and  (20c) are equivalent to
\begin{equation}\tag{32}
 \sum_{i=1, i \neq k}^{K}\left|\mathbf{h}_{k}^{H} \mathbf{w}_{i}\right|^{2}+\sigma_{k}^{2}
=
 \boldsymbol{\psi}^{ {H}}\mathbf{F}  \boldsymbol{\psi}+
 \Re\left\{\boldsymbol{\psi}^{ {H}}\mathbf{J}\right\}+
 \mathbf{E},
\end{equation}
\begin{equation}\tag{33}
 \sum_{i=1}^{K}\left|\mathbf{h}_{k}^{H} \mathbf{w}_{i}\right|^{2}+\sigma_{k}^{2}
= 
 \boldsymbol{\psi}^{ {H}}\mathbf{B}_{2p}  \boldsymbol{\psi}+
 \Re\left\{\boldsymbol{\psi}^{ {H}}\mathbf{J}\right\}+
 \mathbf{A}_{2p},
\end{equation}
and
\begin{equation}\tag{34}
\left\|\mathbf{f}_{k}^{H} \mathbf{\Psi}\right\|^{2} \sigma_{z}^{2}
=
 \boldsymbol{\psi}^{ {H}}\mathbf{D} \boldsymbol{\psi}.
\end{equation}
We define  $\gamma_p= 2^{R_{k}^{\rm min}-r_{c,k}}-1$, and $\gamma_0=\sum_{i=1}^{K} r_{c, i}$, the constraints (20b), (20c) can be respectively reformulated  as

\begin{equation}\tag{35}
   \boldsymbol{\psi}^{  {H}} \mathbf{M}_0 \boldsymbol{\psi}+\Re\left\{\boldsymbol{\psi}^{  {H}} \mathbf{N}_0\right\} \geq \gamma_0\mathbf{A}_{2p}+\mathbf{A}_{1p},
\end{equation}
\begin{equation}\tag{36}
    \boldsymbol{\psi}^{  {H}} \mathbf{M} \boldsymbol{\psi}+\Re\left\{\boldsymbol{\psi}^{  {H}} \mathbf{N}\right\} \geq \gamma_p\mathbf{E}+\mathbf{A}_{1c},
\end{equation}
 where $\mathbf{M}_{0} =\mathbf{B}_{1c}-\gamma_0\mathbf{B}_{2p}-\gamma_0\mathbf{D}$, $\mathbf{N} _{0}=\mathbf{C}_{1c}-\gamma_0\mathbf{C}_{2p}$, $\mathbf{M} =\mathbf{B}_{1p}-\gamma_p\mathbf{F}-\gamma_p\mathbf{D}$, and   $\mathbf{N} =\mathbf{C}_{1p}-\gamma_p\mathbf{J}$.

After the above transformation,  the problem in (20) can be equivalently reformulated as
 \begin{equation}\tag{37}
 \begin{aligned}
    \mathcal{P}_3 :\quad \max_{\boldsymbol{\psi},\boldsymbol{\tau},\boldsymbol{\upsilon}}&\quad\Re\left\{\boldsymbol{\psi}^{ {H}} \mathbf{b}\right\}-\boldsymbol{\psi}^{ {H}} \mathbf{Q}\boldsymbol{\psi}\\
     \text{s.t.} \quad
     &\boldsymbol{\psi}^{  {H}} \mathbf{M} \boldsymbol{\psi}+\Re\left\{\boldsymbol{\psi}^{  {H}} \mathbf{N}\right\} \geq \gamma_p\mathbf{E}+\mathbf{A}_{1p},\\
     &\boldsymbol{\psi}^{  {H}} \mathbf{M}_0 \boldsymbol{\psi}+\Re\left\{\boldsymbol{\psi}^{  {H}} \mathbf{N}_0\right\} \geq \gamma_0\mathbf{A}_{2p}+\mathbf{A}_{1c},\\
     &\boldsymbol{\psi}^{ {H}} \boldsymbol{\Pi} \boldsymbol{\psi} \leq P_{\mathrm {A}}^{\max },
 \end{aligned}
 \end{equation}
 where \begin{equation}\tag{38}
       \mathbf{b}=2 \sum_{k=1}^{K} \sqrt{\left(1+\tau _{k}\right)} \operatorname{diag}\left(\upsilon _{k}^{*} \mathbf{f}_{k}^{ {H}}\right) \mathbf{G} \mathbf{w}_{k}-\sum_{k=1}^{K}\left|\upsilon _{k}\right|^{2} \mathbf{C}_{2p},
 \end{equation}
\begin{equation}\tag{39}
    \mathbf{Q}=\sum_{k=1}^{K}\left|\upsilon _{k}\right|^{2} \mathbf{F}+ \sum_{k=1}^{K}\left|\upsilon _{k}\right|^{2}\mathbf{B}_{2p},
\end{equation}
\begin{equation}\tag{40}
\mathbf{\Pi}=\sum_{i=0}^{K} \operatorname{diag}\left(\mathbf{G} \mathbf{w}_{i}\right)\left(\operatorname{diag}\left(\mathbf{G} \mathbf{w}_{i}\right)\right)^{ {H}}+\sigma_{v}^{2} \mathbf{I}_{L}.
\end{equation}
The optimal solution of $\boldsymbol{\tau}$ and $\boldsymbol{\upsilon}$ can be respectively obtained from (17) and (19). It is obvious that problem $\mathcal{P}_3$ is a standard QCQP problem with respect to $\boldsymbol{\psi}$ and thus the optimal solution of the active RIS precoding matrix $\mathbf{\Psi}$ can be obtained by  CVX  toolbox \cite{cvx}. The iterative algorithm for solving the sum-rate maximization problem in $\mathcal{P}_{0} $ is given in Algorithm 1. 

\begin{algorithm}[h]
\caption{QCQP Based Alternating Algorithm }
\label{al:tra}
\begin{algorithmic}[1]
\STATE Initialize ($\mathbf{\Psi}^{(0)}$, $\mathbf{w}^{(0)}$, $\mathbf{r_c}^{(0)}$). Set iteration number  $t=1$ and convergence tolerance $\epsilon=10^{-5}$.
\REPEAT
\STATE Given $\mathbf{\Psi}^{(t-1)}$,  update ($\mathbf{w}^{(t)}$, $\mathbf{r_c}^{(t)}$) by solving (15).
\STATE Given ($\mathbf{w}^{(t)}$, $\mathbf{r_c}^{(t)}$), update $\mathbf{\Psi}^{(t)}$ by solving (37). 
\STATE Set $t= t+1$.

\UNTIL $\left|R_{tot}^{(t)}-R_{tot}^{(t-1)}\right| \le \epsilon$
\RETURN
 Optimized $\mathbf{\Psi}^{(t)}$, $\mathbf{w}^{(t)}$, and $\mathbf{r_c}^{(t)}$.
\end{algorithmic}
\end{algorithm}
\vspace{-1em}

\subsection{Complexity Analysis}

  The overall computational complexity of Algorithm 1 is mainly determined by the complexity of solving problems (8) and (37). Specifically, the beamforming and rate optimization problem (8) has $5N_TK$ constraints. Since it is solved by the SCA method, the number of iterations of the SCA method is $\mathcal{O}$($\sqrt{5N_TK}\log_2{(1/\epsilon)}$) \cite{cvx}, where  $\epsilon$ is the accuracy of the SCA method.  At each iteration of the SCA method,  the complexity of solving problem (15) is $\mathcal{O}$($S_1^2S_2$), where $S_1=6N_TK$ is the total number of variables and $S_2=6N_TK$ is the total number of constraints. Thus, the computational complexity of the problem (8) is $\mathcal{O}(N_T^{3.5}K^{3.5}\log_2{(1/\epsilon)})$. Considering the  computational complexity of solving the standard QCQP problem (37), the computational complexity of updating $\mathbf{\Psi}$ is $\mathcal{O}\left(\log _{2}(1 / \varepsilon) \sqrt{L+1}(1+2 L) L^{3}\right)$. Hence, the total complexity of Algorithm 1 is $\mathcal{O}(\log_2{(1/\epsilon)}(N_T^{3.5}K^{3.5}+L^{4.5}))$. 
  
  

  \vspace{-0.5em}
\section{Numerical Results}

In this section, we evaluate the performance of the proposed algorithm for the active RIS design. The system settings are summarized in Table \uppercase\expandafter{\romannumeral1}.   Particulaly,  $K$ users are randomly located within a circle with radius of 5 m from the center position. The links are set as Rayleigh fading channel. The pass-loss of the channels is modeled as $P(d)=L_0(d/d_0)^{-\alpha}$, where $d_0=1$ m is the reference distance. 

For comparison, an existing beamforming design with passive RIS-aided RSMA system and an active RIS-aided Space Division Multiple Access (SDMA) system are adopted as baselines.  We  assume that the total power consumption $P^{\rm max}=P_{\rm BS}^{\rm max}+P_{\rm A}^{\rm max}$ is equally allocated to the BS and the active RIS, and the same total power consumption $P^{\rm max}=P_{\rm BS}^{\rm max}$ for the passive RIS system.
\begin{table}[t]
    \centering \caption{Simulation parameters}
    \begin{tabular}{|c|c|}
    \hline
      \textbf{Parameters} & \textbf{Values}
      \\
      \hline
      BS position & (0 m, 0 m) \\
      Active RIS position & (40 m, 30 m)\\
      Users distribution center position & (80 m, 0 m)\\
    
      $K$ & 2\\
      $N_{\rm T}$ & 4\\ 
      Reference distance $d_0$ & 1 m\\
      Path loss $L_0$ & -30 dB\\
 
      Pass-loss exponent of BS to user 1 & 2\\
      Pass-loss exponent of BS to user 2 & 3\\
      Pass-loss exponent of BS to RIS & 3\\
      Pass-loss exponent of RIS to users & 3.5\\
      Noise power $\sigma_z^2 ,  \sigma^2_{k}$ & -80 dBm\\
      Convergence tolerance $\epsilon$ & $10^{-5}$\\
      \hline
    \end{tabular}
\vspace{-1em}
    \label{tab:my_label}
\end{table}
\begin{figure}[!t]
\centering
\includegraphics[width=3.2in]{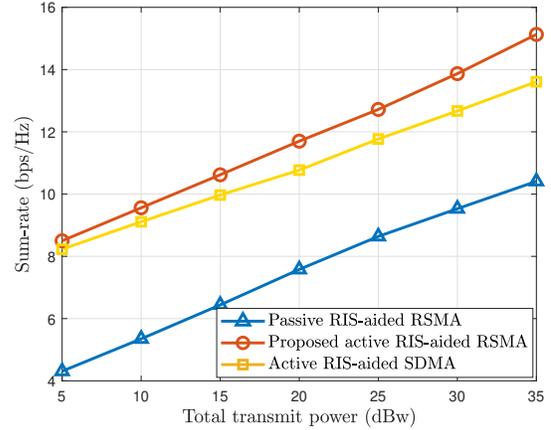}
\caption{Sum-rate versus total power consumption  ($L=16$).}
\vspace{-2em}
\end{figure}

 Fig. 2 shows the sum-rate of different schemes versus the total transmit power. From Fig. 2, we see that, the proposed algorithm achieves outperformance. In particular, RSMA scheme outperforms SDMA scheme, especially under high transmit power. From this figure we also observe that, as the total transmit power  increases, the sum-rate of the active RIS-aided RSMA system increases from 8.34 bps/Hz to 15.68 bps/Hz, while the sum-rate of the passive RIS-aided RSMA system increases from 4.12 bps/Hz to 10.14 bps/Hz. In particular,  the sum-rate of the active RIS increases up to 45\%  compared to the passive RIS.  This is because that  RSMA can further improve the channel capacity by adaptively treating the multi-user interference as noise or interference, while SDMA only treats the multiuser interference as noise.  
  

Fig. 3 illustrates the sum-rate of different schemes versus the number of reflecting elements at the RIS. From Fig. 3,  we can see that the sum-rate of all the schemes monotonically increases with the number of reflecting elements $L$. In particular, when $L$ increases from 8 to 128, the sum-rate of the active RIS-aided RSMA system increases from 9.63 bps/Hz to 12.84 bps/Hz. Moreover,  Fig.~3 demonstrates  that incorporating the active
RIS into RSMA systems leads to a higher sum-rate gain compared to the passive RIS with the same number of reflecting elements,  which indicates that the required value of  $L$ can be greatly reduced when adapting the active RIS. The reason  is  that the signal amplification of active RIS can significantly compensate the performance loss induced by reducing the value of $L$. 

\begin{figure}[!t]
\centering
\includegraphics[width=3.2in]{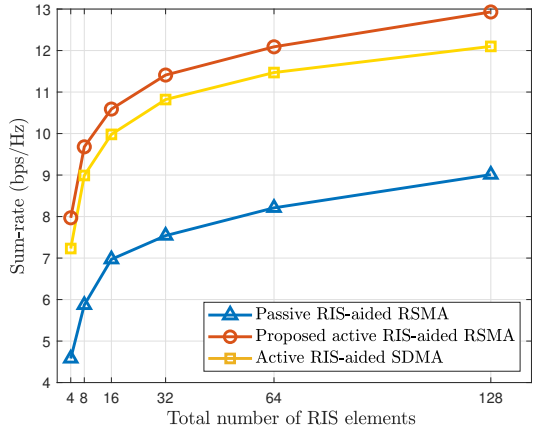}
\caption{Sum-rate versus the number of RIS elements ($P^{\rm max}= 15$ dBw).}
\label{fig:time}
\vspace{-1em}
\end{figure}
\begin{figure}[!t]
\centering
\includegraphics[width=3.2in]{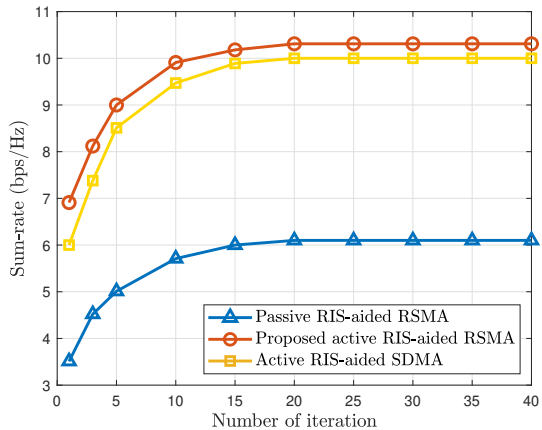}
\caption{Sum-rate versus the number of the iterations.    ($L=16$, $P^{\rm max}= 15$ dBw).}
\label{fig:time}
\vspace{-2em}
\end{figure}
The convergence behavior of the proposed Algorithm 1 is presented in Fig. 4.  From Fig. 4, we observe that  the proposed algorithm can converge within 20 iterations fir active or passive RIS.

\section{Conclusion}
    In this paper, we proposed an active RIS-aided RSMA downlink system. A sum-rate maximization problem is  formulated by jointly optimizing the transmit beamforming vectors, the rate allocation vector, and active RIS precoding matrix. To solve the formulated  problem, an iterative algorithm based on BCD, SCA, FP and QCQP  is proposed, which alternatively optimizes the active RIS precoding matrix and the remaining variables. Numerical results show that the proposed active RIS-aided RSMA system can achieve a higher  sum-rate gain over the passive RIS-aided RSMA system, under the same power consumption.

\bibliographystyle{IEEEtran}
\bibliography{IEEEabrv,ref}
\end{CJK}
\end{document}